\documentclass[12pt]{article}

\usepackage{graphicx}
\usepackage{epsfig}
\usepackage{amsfonts}
\usepackage{caption}
\usepackage{graphicx}
\usepackage{graphicx, epsfig, amssymb} 
\usepackage{bm}
\usepackage{amsfonts}
\usepackage{amsmath}
\usepackage{graphicx}
\usepackage{subcaption}

\usepackage[normalem]{ulem}
\usepackage{xcolor}

\usepackage[colorlinks=true,linktocpage=true,linkcolor=blue,citecolor=blue,urlcolor=blue]{hyperref}

\usepackage{amssymb}
\usepackage{graphicx}
\usepackage{epsfig}
\usepackage{amsfonts}
\usepackage{subcaption}  
\usepackage[utf8]{inputenc}

\date{\today}

\newcommand{\be}{\begin{eqnarray}}
	\newcommand{\ee}{\end{eqnarray}}

\begin{document}
	\title{\bf Resonant electroweak clouds \\ around Schwarzschild black holes}

	\author{
		{\large Carlos Herdeiro}$^{1}$   and    
		{\large Eugen Radu}$^{1}$
		\vspace*{0.2cm}
		\\
		$^{1}${\small Departamento de Matem\'atica da Universidade de Aveiro and } 
		\\ {\small  Centre for Research and Development  in Mathematics and Applications (CIDMA),} 
		\\ {\small    Campus de Santiago, 3810-193 Aveiro, Portugal} 
	}
	
	\date{October 2026}
	
	\maketitle 
	
	\begin{abstract} 
		We construct nonlinear electroweak clouds around Schwarzschild black holes in the bosonic Weinberg-Salam theory. These configurations are supported by harmonically time-dependent charged gauge fields, while their energy-momentum tensor remains static. Regularity at the horizon imposes a resonance condition relating the field frequency to the horizon electrostatic potential. Unlike linear clouds, these solutions exhibit a finite mass and electric charge gap, and exist only up to a maximal horizon size. We obtain both magnetic and electric clouds, including spherically symmetric and dipolar configurations, which reduce to electroweak balls in the vanishing-horizon limit. We also report a new family of dipolar electroweak balls, which can be energetically favoured over their spherical 
	      and magnetic
		counterparts at fixed electric charge. Our results, obtained in the probe limit for electroweak couplings away from their measured values, establish a new realization of nonlinear resonant vector hair in a non-Abelian gauge–Higgs theory.
		
	\end{abstract}
	
	


	\section{Introduction}
	\label{Introduction}
	
	The interplay between black holes (BHs) and non-Abelian
	gauge-Higgs fields offers a particularly interesting arena
	for exploring the limitations of classical no-hair results in the context of potentially realistic high energy physics modeling.
	The bosonic $SU(2)\times U(1)$ Weinberg--Salam (WS) theory
	provides a well-motivated setting for such investigations,
	possessing non-perturbative solutions such as the electroweak
	sphaleron \cite{Klinkhamer:1984di} and its generalizations
	\cite{Kleihaus:1994multi}.
	BHs supporting non-trivial fields in spontaneously
	broken gauge theories have long been known
	\cite{Greene:1992ew}.
	More recently, fully backreacted, magnetically charged BHs
	with electroweak hair have been constructed
	\cite{Gervalle:2024ew,Gervalle:2025ew},
	while nonlinear sphaleron clouds have been found around
	Schwarzschild and Kerr BHs in the probe limit
	\cite{Herdeiro:2026sph}.
	These results illustrate the rich non-perturbative structure
	of electroweak fields in the presence of an event horizon.
	
	A qualitatively different class of non-perturbative
	configurations has recently been reported in
	Ref.~\cite{Herdeiro:2026gbo}
	(see also Ref.~\cite{dosSantosCostaFilho:2026ach}).
	Dubbed \textit{electroweak balls}, these are smooth,
	finite-energy, electrically charged solitons in the bosonic
	WS theory. Their localization mechanism is reminiscent of
	that of $Q$-balls: the charged vector fields possess a
	harmonic time dependence, while the energy-momentum tensor
	remains time independent.
	Unlike sphalerons, these configurations belong to a
	non-topological sector and may be regarded as electroweak
	counterparts of Proca-Higgs balls
	\cite{Herdeiro:2023lze}, with the vector-boson masses
	generated by the Higgs mechanism.
	This connection is particularly suggestive in view of the
	rich phenomenology of Proca stars, including their
	non-spherical ground state \cite{Herdeiro:2023wqf},
	dynamically robust spinning configurations
	\cite{Sanchis-Gual:2019ljs}, and the formation of BHs
	with Proca hair through superradiance
	\cite{East:2017ovw}.
	
	A natural question is whether electroweak balls can
	accommodate an event horizon while preserving their
	stationary energy-momentum tensor.
	The answer is not obvious, given the restrictions on
	static BH configurations supported by harmonically
	time-dependent scalar and Proca fields
	\cite{Pena:1997cy,Herdeiro:2016tmi}.
	A possible mechanism is provided by \textit{resonance}:
	the field frequency is synchronized with the electrostatic
	potential at the horizon, allowing a regular configuration
	without matter flux across it.
	Such a mechanism has been demonstrated for nonlinear,
	electrically charged scalar clouds around static BHs
	\cite{Herdeiro:2020xmb}.
	Whether it can be realized by the charged vector fields
	of a non-Abelian gauge--Higgs theory is the central
	question addressed here.
	
	In this work, we answer this question affirmatively by
	constructing \textit{nonlinear electroweak clouds} on a
	fixed Schwarzschild background.
	The solutions satisfy a resonance condition relating the
	harmonic frequency of the charged gauge fields to the
	horizon electrostatic potential.
	We obtain both magnetic- and electric-type configurations,
	including spherically symmetric and axially symmetric
	families.
	The clouds possess a finite mass and electric charge gap
	along the families investigated, with no connection to
	linear clouds, and exist only up to a maximal horizon size.
	In the vanishing horizon limit, they continuously connect
	to electroweak balls.
	Thus, our construction provides a realization of nonlinear
	resonant vector clouds within the WS gauge-Higgs theory,
	distinct from the previously known magnetic electroweak
	BH hair and sphaleron clouds.
	
	A further result concerns the solitonic sector itself.
	Motivated by the electric dipolar Proca stars of
	Ref.~\cite{Herdeiro:2023wqf}, we construct a new family
	of axially symmetric, electric dipolar electroweak balls.
	These configurations may have lower energy than their
	spherically symmetric and magnetic counterparts at fixed electric
	charge.
They also admit resonant generalizations in
	the Schwarzschild background, exhibiting the same
	qualitative branch structure as the other cloud families.
	
	Our analysis is performed in the probe approximation,
	neglecting the gravitational backreaction of the
	electroweak fields.
	Moreover, as for the electroweak balls of
	Ref.~\cite{Herdeiro:2026gbo}, the solutions constructed
	here occur for gauge and Higgs couplings away from their
	measured Standard Model values.
	They should therefore be viewed as solutions of
	electroweak-type gauge-Higgs theories rather than
	established configurations at the physical Standard
	Model parameters.
	
	The paper is organized as follows.
	In Sec.~\ref{section:model} we briefly review the bosonic
	WS theory and the field ans\"atze.
	Section~\ref{section:balls} presents the new electric
	dipolar electroweak balls.
	In Sec.~\ref{section:clouds} we construct the resonant
	electroweak clouds on Schwarzschild backgrounds and
	discuss their main properties.
	We conclude in Sec.~\ref{section:conclusions}.

	\section{The bosonic Weinberg-Salam model}
	\label{section:model}
	
	We consider the bosonic sector of the WS
	theory on a prescribed spacetime geometry. The model
	combines non-Abelian gauge interactions with the Higgs
	mechanism, giving rise to massive charged vector bosons
	while retaining an unbroken electromagnetic $U(1)$
	symmetry. These features provide the essential ingredients
	for the localized configurations considered below, both
	in Minkowski spacetime and in the presence of a BH horizon.
	
	The field content consists of an $SU(2)$ gauge connection
	$V_\mu^a$ ($a=1,2,3$), a $U(1)$ gauge potential $a_\mu$,
	and a complex Higgs doublet $\Phi$. With the conventions
	adopted here, the action is
	\begin{equation}
		\label{SM}
		S_{\rm WS}=-\frac{1}{4\pi}
		\int d^4x\,\sqrt{-g}\,
		\left[
		\frac{1}{4}F_{\mu\nu}^aF^{a\mu\nu}
		+\frac{1}{4}f_{\mu\nu}f^{\mu\nu}
		+(D_\mu\Phi)^\dagger D^\mu\Phi
		+\lambda\left(\Phi^\dagger\Phi-\frac{v^2}{2}\right)^2
		\right].
	\end{equation}
	The corresponding field strengths and the covariant
	derivative of the Higgs doublet are defined by
	\begin{align}
		F_{\mu\nu}^a &=
		\partial_\mu V_\nu^a-\partial_\nu V_\mu^a
		+g\epsilon^{abc}V_\mu^bV_\nu^c,
		\nonumber\\
		f_{\mu\nu} &=
		\partial_\mu a_\nu-\partial_\nu a_\mu,
		\nonumber\\
		D_\mu\Phi &=
		\left(
		\partial_\mu-\frac{ig}{2}\tau^aV_\mu^a
		-\frac{ig'}{2}a_\mu
		\right)\Phi ,
	\end{align}
	where $\tau^a$ are the Pauli matrices, $g$ and $g'$
	are the gauge couplings, and $\lambda$ and $v$
	specify the Higgs potential.
	
	The symmetry breaking vacuum can be represented,
	in a suitable gauge, by
	\begin{equation}
		\label{H01}
		\langle\Phi\rangle=
		\frac{v}{\sqrt{2}}
		\begin{pmatrix}
			0\\1
		\end{pmatrix}.
	\end{equation}
	The resulting spectrum contains charged massive vector
	bosons $W^\pm_\mu=(V_\mu^1\mp iV_\mu^2)/\sqrt{2}$,
	a massive neutral vector boson $Z_\mu$, a massless
	photon, and the Higgs scalar. {Their masses are
		%
		$
		M_W= {gv}/{2},
		$
		$
		M_Z=\frac{v}{2}\sqrt{g^2+g'^2},
		$
		and 
		$
		M_H=v\sqrt{2\lambda}
		$,
		respectively.
	}
	The electroweak mixing angle and the elementary
	electromagnetic coupling satisfy
	\begin{equation}
		\tan\theta_{\rm W}=\frac{g'}{g},
		\qquad
		e=g\sin\theta_{\rm W}.
	\end{equation}
	In a gauge where the Higgs doublet approaches
	(\ref{H01}) asymptotically, the neutral gauge fields
	are conveniently combined into
	\begin{equation}
		\label{Zemfield}
		Z_\mu=\cos\theta_{\rm W}V_\mu^3
		-\sin\theta_{\rm W}a_\mu,
		\qquad
		\tilde A_\mu=\sin\theta_{\rm W}V_\mu^3
		+\cos\theta_{\rm W}a_\mu.
	\end{equation}
	The latter identifies the asymptotic electromagnetic
	potential, with field strength
	$\tilde F_{\mu\nu}=\partial_\mu\tilde A_\nu
	-\partial_\nu\tilde A_\mu$.
	In particular, it allows us to extract the electric
	charge from the asymptotic electromagnetic flux.
	
	Variation of the action (\ref{SM}) yields the coupled
	Yang-Mills, Abelian gauge and Higgs equations,
	\begin{equation}
		\label{eqsSM}
		\begin{aligned}
			D_\mu F^{a\mu\nu}
			&=\frac{ig}{2}
			\left[
			\Phi^\dagger\tau^aD^\nu\Phi
			-(D^\nu\Phi)^\dagger\tau^a\Phi
			\right],
			\\[2mm]
			\nabla_\mu f^{\mu\nu}
			&=\frac{ig'}{2}
			\left[
			\Phi^\dagger D^\nu\Phi
			-(D^\nu\Phi)^\dagger\Phi
			\right],
			\\[2mm]
			D_\mu D^\mu\Phi
			&=2\lambda
			\left(\Phi^\dagger\Phi-\frac{v^2}{2}\right)\Phi ,
		\end{aligned}
	\end{equation}
	where $D_\mu$ denotes the appropriate gauge covariant
	derivative, acting in the adjoint representation in
	the first equation.
	
	A central feature of this system is that the charged
	$W^\pm$ fields can carry a harmonic time dependence
	without inducing a time dependence in the gauge invariant
	observables, including the energy-momentum tensor.
	Their coupling to the neutral electric potentials also
	provides the possibility of satisfying a resonance
	condition at a BH horizon. In the following, we exploit
	these properties to construct electroweak balls and their
	nonlinear cloud counterparts.

	\section{Electroweak balls}
	\label{section:balls}
	
	We first consider localized solutions of the WS theory
	in a Minkowski spacetime background, focusing on the
	electric sector.
	After motivating the classification of the field
	configurations in terms of their Proca counterparts,
	we extend the ans\"atze of~\cite{Herdeiro:2026gbo} and construct a new family
	of axially symmetric, electric dipolar electroweak balls.
	These solutions will also provide the solitonic limit
	of the corresponding BH clouds discussed in
	Sec.~\ref{section:clouds}.

	\subsection{A Proca-inspired classification}
	
	A useful guideline for constructing electroweak balls
	is provided by the multipolar classification of Proca
	stars developed in Ref.~\cite{Herdeiro:2026agu},
	following analogous ideas for scalar boson stars
	\cite{Herdeiro:2020kvf}.
	In this approach, the solitonic configurations are
	interpreted as nonlinear, regular counterparts of
	the multipolar modes of a massive vector field.
	They are characterized by a multipole index
	$\ell\geq 0$ and organized into two classes,
	denoted by \textbf{(e)} and \textbf{(m)},
	according to their electric- or magnetic-type
	vector structure.
	
	For a complex Proca field with harmonic time dependence,
	the magnetic-type ansatz takes the simple form
	\begin{equation}
		\label{ansatzM}
		\mathcal{A}
		=e^{-i\omega t}S(r,\theta)\,d\varphi ,
	\end{equation}
	whereas the electric-type ansatz involves both temporal
	and spatial components,
	\begin{equation}
		\label{ansatzE}
		\mathcal{A}
		=e^{-i\omega t}
		\left[
		iH_0(r,\theta)\,dt
		+\frac{H_1(r,\theta)}{r}\,dr
		+H_2(r,\theta)\,d\theta
		\right].
	\end{equation}
	Here $\omega>0$ is the field frequency, and
	$(r,\theta,\varphi)$ are spherical coordinates.
	For the electric sector, the multipolar character
	is encoded, in particular, in the angular dependence
	of the temporal amplitude $H_0$.
	The familiar spherically symmetric Proca stars
	\cite{Brito:2015pxa} correspond to the
	$\ell=0$ electric sector, while higher multipoles
	include electric dipoles ($\ell=1$) and quadrupoles
	($\ell=2$) \cite{Herdeiro:2023wqf}.
	
	The regularization of these multipolar configurations
	can be achieved through their nonlinear coupling to
	gravity, yielding self-gravitating Proca stars.
	Alternatively, localized solutions can arise in
	flat spacetime through interactions with a Higgs
	field, as illustrated by the Proca-Higgs model
	\cite{Herdeiro:2023lze}.
	The latter mechanism is particularly relevant for
	the WS theory, where the charged vector bosons
	acquire their mass through spontaneous symmetry
	breaking.
	
	Indeed, the electric electroweak balls constructed
	in~\cite{Herdeiro:2026gbo} may be interpreted
	as counterparts of the $\ell=0$ electric Proca
	configurations.
	This analogy suggests the existence of higher
	multipolar electroweak balls, despite the additional
	constraints imposed by the non-Abelian gauge
	structure and the coupling to the electromagnetic
	and Higgs sectors.
	In what follows, we establish numerically the
	existence of the simplest such extension,
	corresponding to electric dipolar configurations.
	
	Motivated by this correspondence, we retain the
	terminology \textbf{(e)} and \textbf{(m)} for the
	electroweak ans\"atze.
	We emphasize, however, that this classification
	refers to the structure of the charged,
	harmonically time-dependent gauge fields, and
	should not be confused with a classification
	of the gauge-invariant electromagnetic field
	into electric and magnetic multipoles.
	
	For both sectors, we adopt a purely electric
	$U(1)$ gauge potential and a real Higgs amplitude,
	\begin{equation}
		\label{U1H}
		a_\mu dx^\mu
		=\frac{1}{g'}a_0(r,\theta)\,dt,
		\qquad
		\Phi
		=\frac{v}{\sqrt{2}}\phi(r,\theta)
		\begin{pmatrix}
			0\\
			1
		\end{pmatrix}.
	\end{equation}
	The functions $a_0$ and $\phi$ are independent
	of time, with $\phi\to 1$ asymptotically.
	The distinction between the electric- and
	magnetic-type configurations is then implemented
	through two different ans\"atze for the
	non-Abelian $SU(2)$ gauge potential, which
	we specify below.

	
	The magnetic Ansatz has a simple  expression, in terms of two potentials
	$S$ and $A_0$  \cite{Herdeiro:2026gbo}:
	\begin{eqnarray}
		\label{YM-ansatz-m}
	{{\bf m}\text{-Ansatz:}}\qquad	V_\mu dx^\mu
		=
		S(r,\theta)
		\left(
		\sin\omega t\,\frac{\tau_1}{g}
		+
		\cos\omega t\,\frac{\tau_2}{g}
		\right)d\varphi
		+
		A_0(r,\theta)\frac{\tau_3}{g}\,dt .
	\end{eqnarray}
	
	The resulting equations and boundary conditions are presented in Ref. \cite{Herdeiro:2026gbo},
	together with a discussion of the basic properties
	of the solutions.
	
	In what follows we shall focus on the electric sector, where we report on the existence of a new class of solutions.
	
	\subsection{Electric-type electroweak balls}
	\label{sec:electric-balls}
	
	The electric sector of~\cite{Herdeiro:2026gbo} contains the
	spherically symmetric electroweak balls. We now extend its field ansatz
	to axial symmetry, focusing on the first non-spherical configurations:
	electric dipoles. Besides enlarging the solitonic solution space, these
	configurations provide the zero-horizon limit of the dipolar electroweak
	clouds discussed in Sec.~\ref{section:clouds}.
	
	\subsubsection{Electric ansatz and reduced equations}
	\label{sec:ansatz}
	
	We parameterize the $SU(2)$ connection by four real functions
	$H_0,H_1,H_2,A_0$ of $(r,\theta)$. In the matrix representation
	$V_\mu=V_\mu^a\tau_a$, the electric-type ansatz reads
	\begin{equation}
		\label{YM-ansatz}
	{{\bf e}\text{-Ansatz:}}\qquad	V_\mu dx^\mu=\frac{1}{g}\left[
		\mathsf P(\omega t)\left(\frac{H_1}{r}\,dr+H_2\,d\theta\right)
		+\mathsf T(r,\theta,t)\,dt\right],
	\end{equation}
	where
	\begin{equation}
		\begin{gathered}
			\mathsf P(\omega t)=
			\begin{pmatrix}
				0&e^{i\omega t}\\
				e^{-i\omega t}&0
			\end{pmatrix}, \qquad 
			\mathsf T(r,\theta,t)=
			\begin{pmatrix}
				A_0&-iH_0e^{i\omega t}\\
				iH_0e^{-i\omega t}&-A_0
			\end{pmatrix}.
		\end{gathered}
	\end{equation}
	Equivalently, in the Pauli-matrix basis,
	\begin{equation}
		\label{YM-ansatz-v}
		\begin{aligned}
			V_\mu dx^\mu=\frac1g\Big[
			\left(\frac{H_1}{r}\,dr+H_2\,d\theta\right)
			(\cos\omega t\,\tau_1-\sin\omega t\,\tau_2)
			+\left(H_0(\sin\omega t\,\tau_1+\cos\omega t\,\tau_2)
			+A_0\tau_3\right)dt\Big].
		\end{aligned}
	\end{equation}
	The $U(1)$ and Higgs fields are as in Eq.~\eqref{U1H}.
	The vacuum corresponds to $H_i=A_0=a_0=0$ and $\phi=1$.
	
	Although the charged gauge potentials oscillate with frequency
	$\omega$, this time dependence is a gauge rotation; all gauge-invariant
	observables, including the energy-momentum tensor, are time independent.
	Indeed,
	\begin{equation}
		\label{gauge_rotation}
		\partial_t V_\mu^a=\mathcal D_\mu\Psi^a,
		\qquad \Psi^a=\frac{\omega}{g}\delta^a_3,
	\end{equation}
	where $\mathcal D_\mu$ is the $SU(2)$ covariant derivative in the
	adjoint representation. Introducing
	$\mathcal T^a\equiv V_t^a-\Psi^a$, one obtains
	\begin{equation}
		\label{relation1}
		F_{\mu t}^a=\mathcal D_\mu\mathcal T^a.
	\end{equation}

	Substitution of Eqs.~\eqref{YM-ansatz} and \eqref{U1H} into the WS
	action gives, after removing an irrelevant overall factor and the time
	integral,
	\begin{equation}
		\label{Seffn}
		\mathcal S_{\rm eff}=\int dr\,d\theta\,r^2\sin\theta\,
		\bigl(\mathcal L_{\rm YM}+\mathcal L_{\rm M}+\mathcal L_{\rm H}\bigr),
	\end{equation}
	with
	\begin{align}
		\mathcal L_{\rm YM}&=\frac{1}{2g^2}\bigg\{
		\frac{1}{r^2}\left(\frac{H_{1,\theta}}r-H_{2,r}\right)^2
		-\left[H_{0,r}+(\omega-A_0)\frac{H_1}{r}\right]^2
		-\frac{1}{r^2}\left[H_{0,\theta}+(\omega-A_0)H_2\right]^2
		\nonumber\\[-1mm]
		&\quad-\left(A_{0,r}+\frac{H_0H_1}{r}\right)^2
		-\frac{1}{r^2}\left(A_{0,\theta}+H_0H_2\right)^2
		\bigg\},
		\nonumber\\[1mm]
		\mathcal L_{\rm M}&=-\frac{1}{2g'^2}
		\left(a_{0,r}^2+\frac{a_{0,\theta}^2}{r^2}\right),
		\nonumber\\[1mm]
		\mathcal L_{\rm H}&=\frac{v^2}{2}\bigg\{
		\phi_{,r}^2+\frac{\phi_{,\theta}^2}{r^2}
		+\frac{\phi^2}{4}\left[
		\frac{H_1^2+H_2^2}{r^2}-H_0^2-(a_0-A_0)^2\right]
		+\frac{\lambda v^2}{2}(1-\phi^2)^2
		\bigg\}.
		\label{reduced-L}
	\end{align}
	Here and below, subscripts following a comma denote partial
	derivatives. The equations obtained by varying the reduced action are
	consistent with the corresponding components of the covariant equations
	\eqref{eqsSM}. The latter also yield the relation
	\begin{equation}
		\label{constraint}
		\begin{aligned}
			&2rH_1\phi_{,r}+2H_2\phi_{,\theta}+\bigl(rH_{1,r}+H_{2,\theta}+H_1+H_2\cot\theta\bigr)\phi-r^2(\omega-a_0)H_0\phi=0,
		\end{aligned}
	\end{equation}
	which is a differential consequence of the remaining field equations
	and provides a useful consistency check.
	
	The reduced system is invariant under the simultaneous constant shift
	\begin{equation}
		\label{gauge-shift-e}
		A_0\longrightarrow A_0+\Omega,\qquad
		a_0\longrightarrow a_0+\Omega,\qquad
		\omega\longrightarrow\omega+\Omega.
	\end{equation}
	We fix this residual gauge freedom by requiring
	$A_0,a_0\to0$ at spatial infinity. The frequency used below is defined
	with respect to this gauge.
	
	The spherical solutions of Ref.~\cite{Herdeiro:2026gbo} are recovered
	by taking
	\begin{equation}
		\label{spherical}
		H_1=rH(r),\qquad H_2=0,\qquad H_0=V(r),
	\end{equation}
	with $A_0,a_0$ and $\phi$ depending on $r$ only (up to the
	frequency sign convention of that reference).
	
	\subsubsection{Dipolar boundary conditions and global quantities}
	\label{sec:BCs}
	
	For regular solitons, the origin and asymptotic boundary conditions are
	\begin{align}
		&H_i\big|_{r=0}=0,\qquad
		\partial_r(A_0,a_0,\phi)\big|_{r=0}=0,
		\label{bc-origin}\\
		&H_i,A_0,a_0\longrightarrow0,\qquad
		\phi\longrightarrow1\quad(r\to\infty).
		\label{bc-infty}
	\end{align}
	On the symmetry axis, $\theta=0,\pi$, regularity requires
	\begin{equation}
		\label{bc-axis}
		\partial_\theta(H_0,H_1,A_0,a_0,\phi)=0,
		\qquad H_2=0.
	\end{equation}
	For the electric dipole, $H_0$ and $H_1$ are odd under the equatorial
	reflection $\theta\to\pi-\theta$, whereas $H_2,A_0,a_0$ and $\phi$
	are even. Thus the numerical domain may be restricted to
	$0\leq\theta\leq\pi/2$, with
	\begin{equation}
		\label{eq.boundary-Pi2}
		\begin{gathered}
			H_0=H_1=0,\\
			\partial_\theta(H_2,A_0,a_0,\phi)=0
			\qquad(\theta=\pi/2).
		\end{gathered}
	\end{equation}

	The dipolar character is manifest in the far-field angular dependence
	\begin{equation}
		\label{rinf-e}
		H_0(r,\theta) \simeq h_0(r)\cos\theta,\quad
		H_1(r,\theta) \simeq h_1(r)\cos\theta,\quad
		H_2(r,\theta) \simeq h_2(r)\sin\theta.
	\end{equation}
	In particular, $h_1$ and $h_2$ are independent radial amplitudes,
	related asymptotically by Eq.~\eqref{constraint}. Neglecting the
	subleading Coulomb and Higgs-gradient corrections, this gives
	\begin{equation}
		\label{dipole-radial-relation}
		2h_2\simeq \omega r^2h_0-h_1-rh_1'.
	\end{equation}
	The charged vector fields have a characteristic inverse localization
	length $\kappa=\sqrt{M_W^2-\omega^2}$, implying the necessary bound-state
	condition
	\begin{equation}
		\label{frequency-bound}
		\omega\leq M_W.
	\end{equation}
	The neutral electrostatic fields have a common Coulomb tail,
	\begin{equation}
		\label{coulomb-tail}
		A_0=\frac{Q}{r}+o(r^{-1}),\qquad
		a_0=\frac{Q}{r}+o(r^{-1}),\qquad
		\phi\longrightarrow1.
	\end{equation}
	The homogeneous massive neutral and Higgs modes decay on scales
	$M_Z^{-1}$ and $M_H^{-1}$, respectively. In general, the full
	subleading asymptotics also receive contributions sourced by the
	charged fields; the leading Coulomb term is sufficient for extracting
	the electric charge.
	
	The total energy $M$ is obtained by integrating the energy density
	$\rho=-T^t{}_t$ over the flat spatial slice. The electric charge,
	defined by the electromagnetic flux at spatial infinity, is
	\begin{equation}
		\label{Qel}
		\begin{aligned}
			\mathcal Q&=\frac{1}{4\pi}\oint_{S^2_\infty}{}^\star
			\!\widetilde F_{\theta\varphi}\,d\theta\,d\varphi=\frac{\sin\theta_{\rm W}}{g}Q
			+\frac{\cos\theta_{\rm W}}{g'}Q
			=\frac{Q}{e}.
		\end{aligned}
	\end{equation}

	A useful integral identity follows from the field equations together
	with Eq.~\eqref{relation1}. Here $a_t=a_0/g'$ denotes the Abelian
	time component:
	\begin{equation}
		\label{Me1}
		\begin{aligned}
			&\frac12 F_{\mu t}^aF^{a\mu t}
			+\frac12 f_{\mu t}f^{\mu t}
			+(D_t\Phi)^\dagger D^t\Phi
			=\frac{1}{\sqrt{-g}}\partial_\mu\left\{
			\sqrt{-g}\left[
			\frac12\mathcal T^aF^{a\mu t}
			+\frac12\left(a_t-\frac{\omega}{g'}\right)f^{\mu t}
			\right]\right\}.
		\end{aligned}
	\end{equation}
	For globally regular solitons, the origin contributes no boundary
	term. Consequently, the electric and time-covariant contribution to the
	energy can be expressed entirely in terms of the asymptotic data:
	\begin{equation}
		\label{Me1a}
		\begin{aligned}
			M_{\rm el}
			&=- \frac{1}{4 \pi}\lim_{r\to\infty}\oint_{S^2_r} 
			\Bigg[\frac12\mathcal T^aF^{art}+\frac12\left(a_t-\frac{\omega}{g'}\right)f^{rt}
			\Bigg]r^2\,d\Omega= \frac{\omega \mathcal Q}{2e} ,
		\end{aligned}
	\end{equation}
	where $d\Omega=\sin\theta\,d\theta\,d\varphi$.
	The left-hand side of Eq.~\eqref{Me1} is non-positive in
	Minkowski spacetime. Thus a vanishing product
	$\omega\mathcal Q$ forces the electric/time-covariant contributions
	to vanish. 
{Then one can prove that no 
axisymmetric solitonic configurations  can exist in this case.}

	\begin{figure*}[t]
		\centering
		\includegraphics[width=0.48\textwidth]{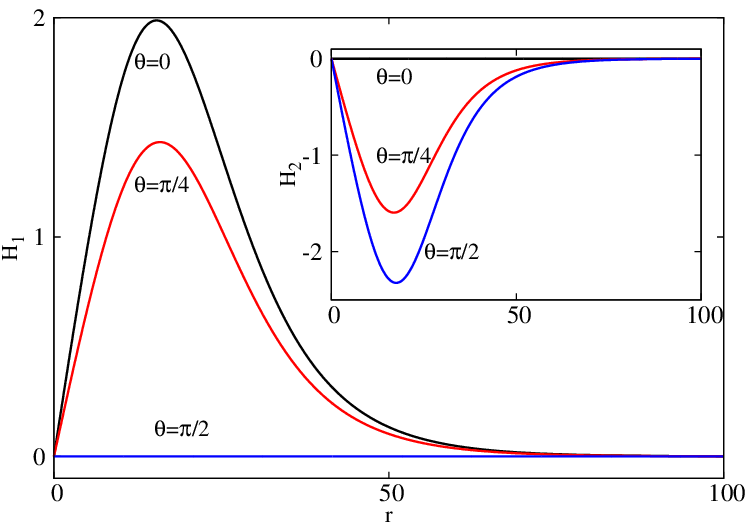}\hfill
		\includegraphics[width=0.48\textwidth]{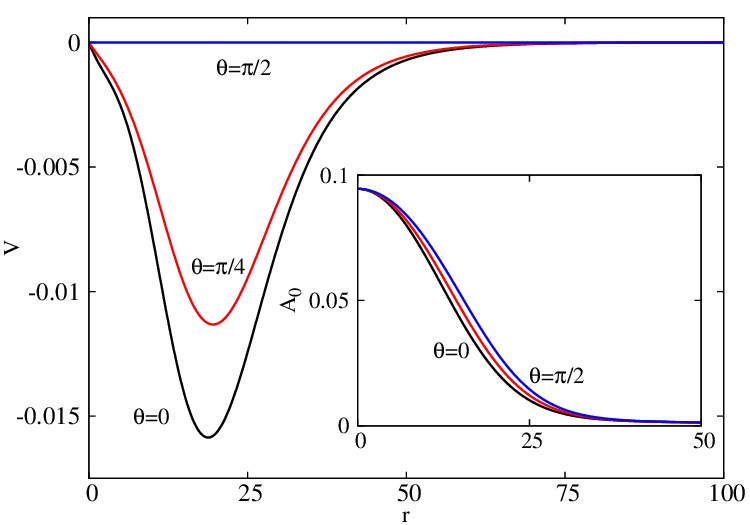}\\[1mm]
		\includegraphics[width=0.48\textwidth]{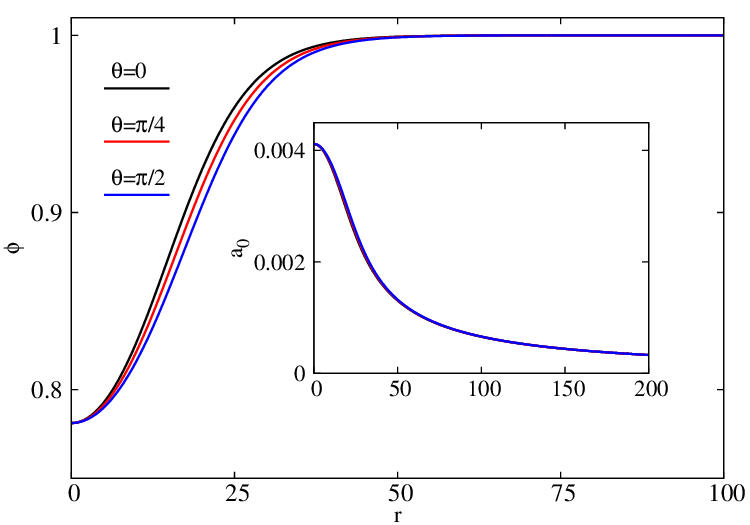}\hfill
		\includegraphics[width=0.48\textwidth]{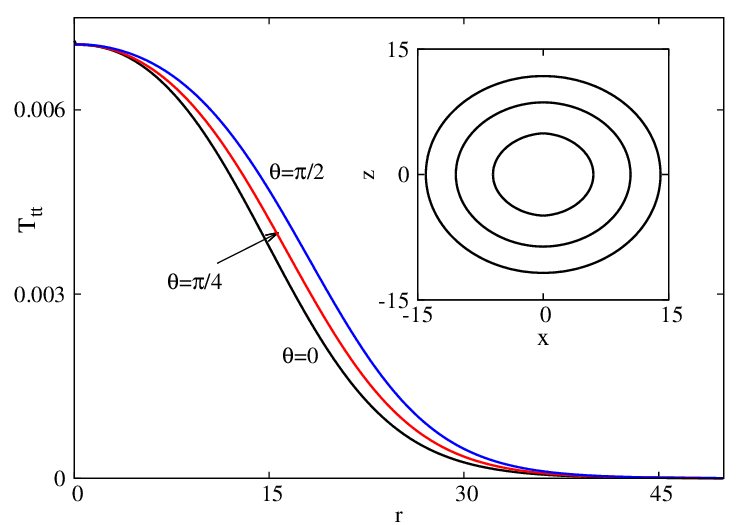}
        \captionsetup{font=small}
		\caption{Profiles of a representative electric dipolar electroweak
			ball with $\omega/M_W=0.98$, $g'/g=0.04$ and $\beta=0.10955$,
			corresponding to the marked configuration in
			Fig.~\ref{wMQ} (left). The radial profiles are shown for several
			polar angles; the inset in the energy-density panel displays
			isodensity contours in the $(x,z)$ plane.}
		\label{axial-profile}
	\end{figure*}
    
	\subsubsection{Numerical solutions and physical properties}
	\label{results}
	\label{sec:balls-numerics}
	
	For the numerical treatment we introduce the dimensionless coordinate
	$\bar r=gv\,r$, and rescale
	$H_0=gv\,\bar H_0$, $A_0=gv\,\bar A_0$,
	$a_0=gv\,\bar a_0$ and $\omega=gv\,\bar\omega$.
	The problem is then controlled by $g'/g$, the Higgs coupling
	parameter
	\begin{equation}
		\label{beta-def}
		\beta^2\equiv\frac{\lambda}{g^2}
		=\frac18\left(\frac{M_H}{M_W}\right)^2,
	\end{equation}
	and the frequency. For orientation, the measured electroweak values
	are $g'/g\simeq0.54880$ and $\beta^2\simeq0.3029$.
	We express the energy and electric charge in units of
	$M_0=v/g$ and $\mathcal Q_0=1/e$, respectively, suppressing these
	units in the figures. Frequencies displayed in the figures are
	normalized by $M_W$.
	
	The field equations are solved with the finite-difference package of
	Ref.~\cite{schoen}, using a nonuniform grid and adjustable
	consistency order. The estimated numerical error {for the solutions in this work}
    is typically below $10^{-4}$. 
    The electric dipolar configurations are obtained by
	continuation from suitable Proca-type profiles, initially taking
	$A_0=a_0=0$ and $\phi=1$, and subsequently solving the coupled WS
	system.
	
	A representative solution is shown in Fig.~\ref{axial-profile}.
	The charged-vector profiles are regular and localized, and the Higgs
	amplitude remains nonzero throughout the numerical domain. The energy density is
	maximal near the origin; its isodensity surfaces are mildly oblate,
	in contrast with the prolate morphology of the Proca configurations
	used to motivate the construction.

\begin{figure*}[t]
		\centering
		\includegraphics[width=0.48\textwidth]{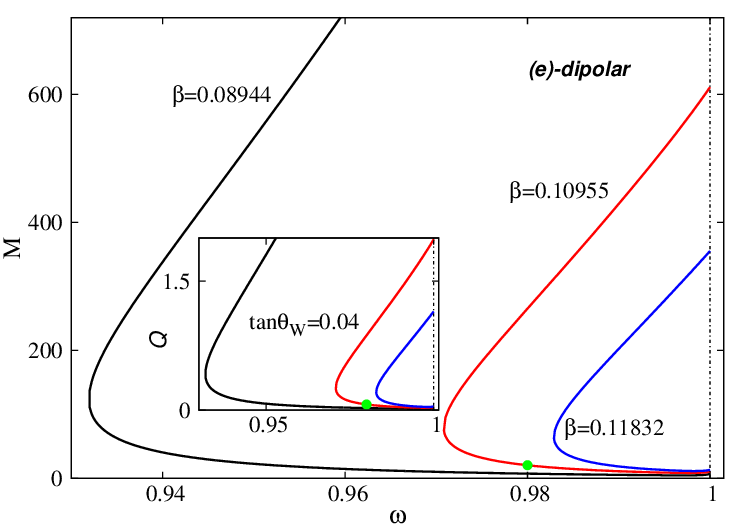}\hfill 
				\includegraphics[width=0.48\textwidth]{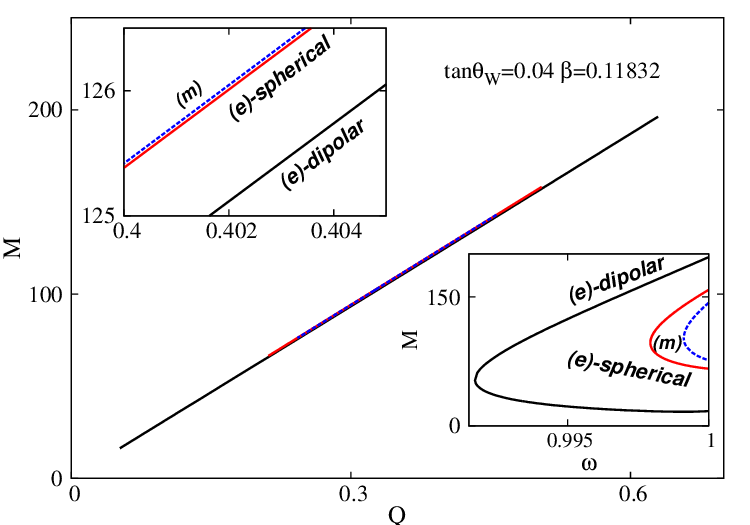}
        \captionsetup{font=small}
		\caption{\textit{Left:} Total energy and electric charge versus
			normalized frequency for representative families of electric
			dipolar electroweak balls. \textit{Right:} Energy versus electric charge
			for the spherical,  dipolar and magnetic families. 
            {The upper inset provides a close-up of the energy–charge curves to highlight their ordering, and the lower inset shows the mass as a function of frequency for all three families.}
		}
		\label{wMQ}
	\end{figure*}

	As in the spherical case \cite{Herdeiro:2026gbo}, the solutions occur
	only in a restricted frequency interval,
	$\omega_{\min}<\omega\leq M_W$, with a nonzero mass gap.
	For the parameter values explored, varying $\omega$ yields two
	branches that meet at $\omega_{\min}$; the mass and electric charge
	along these branches are illustrated in Fig.~\ref{wMQ} (left).
	In particular, the configurations do not bifurcate from the trivial
	vacuum in the families investigated.
	
	{A distinctive feature is the energetic comparison with the other known  
		EW balls \cite{Herdeiro:2026gbo}.
For the couplings shown in Fig.~\ref{wMQ} (right panel), the electric dipolar balls have lower energy than the electric spherical and magnetic-type balls at the same charge over the range where all three are shown.
}
	This comparison
	suggests that the energetically preferred morphology need not be
	spherical; it does not, by itself, establish dynamical stability.
	
	Finally, as for the spherical electroweak balls, no dipolar solutions
	have so far been found at the measured values of both $g'/g$ and
	$\beta^2$. In the explored families the physical mixing-angle
	parameter can be approached for sufficiently small Higgs coupling,
	but the measured Higgs coupling remains outside the region in which
	solutions have been obtained.

	\section{Electroweak clouds on Schwarzschild backgrounds}
	\label{section:clouds}
	
	We now investigate whether the electroweak balls discussed above can
	coexist with a BH horizon. The harmonic time dependence of the charged
	vector fields makes this extension nontrivial in view of the
	restrictions on stationary matter at static horizons
	\cite{Pena:1997cy,Herdeiro:2016tmi}: regular configurations must
	be compatible with the absence of flux through the horizon.
	As we shall see, this is achieved through a resonance
	condition fixing the horizon electrostatic potential in terms of the
	field frequency.
	
	\subsection{Probe limit and horizon boundary conditions}
	
	We solve the matter equations (\ref{eqsSM}) on a fixed Schwarzschild
	background, neglecting the gravitational backreaction of the electroweak
	fields. In isotropic coordinates, the line element reads
	\begin{equation}
		\label{schw}
		\begin{aligned}
			ds^2={}&-\left(\frac{1-r_H/r}{1+r_H/r}\right)^2dt^2+\left(1+\frac{r_H}{r}\right)^4
			\left[dr^2+r^2d\Omega_2^2\right],
		\end{aligned}
	\end{equation}
	where $d\Omega_2^2=d\theta^2+\sin^2\theta\,d\varphi^2$.
	The event horizon is located at $r=r_H>0$ and has area
	$A_H=64\pi r_H^2$. The resulting matter configurations will be termed
	\emph{electroweak clouds}. The probe approximation isolates the
	nonlinear gauge--Higgs mechanism from gravitational self-interactions;
	for the measured electroweak scale, the dimensionless gravitational
	coupling $\alpha^2=Gv^2\sim 10^{-34}$ is very small
	\cite{Herdeiro:2026sph}.
	
	For electric-type clouds we retain the ans\"atze
	(\ref{U1H}) and (\ref{YM-ansatz}), now defined on the exterior region
	$r\geq r_H$. At spatial infinity and on the symmetry axis (and,
	for dipolar solutions, the equatorial plane), we impose the conditions
	specified for the corresponding electroweak balls. Regularity at the
	horizon leads to
	\begin{equation}
		\label{BCh}
		\left.
		\begin{gathered}
			H_1=H_0=0,\qquad \partial_r H_2=\partial_r\phi=0,\\
			A_0=a_0=\omega
		\end{gathered}
		\right\}\qquad (r=r_H).
	\end{equation}
	These conditions apply to both spherical and dipolar electric clouds.
	
	%
	%
	%
	%

	We also consider the magnetic-type sector of
	Ref.~\cite{Herdeiro:2026gbo}.  
	{ Its non-Abelian gauge potential is given by (\ref{YM-ansatz-m}) }
	%
	%
	whereas the Abelian potential and Higgs doublet retain the form
	(\ref{U1H}). 
	{In this case, the boundary   conditions at the  horizon are 
		%
		$
		\partial_r S=\partial_r\phi=0,
		$
		$
		A_0=a_0=\omega.
		$
		%
	}
	At infinity we require $S=A_0=a_0=0$ and $\phi=1$; on the symmetry
	axis, $S=0$ and
	$\partial_\theta A_0=\partial_\theta a_0=
	\partial_\theta\phi=0$.
	
	The common condition $A_0(r_H)=a_0(r_H)=\omega$ is the key physical
	ingredient. Indeed, Eq.~(\ref{Zemfield}) and our conventions imply
	\begin{equation}
		\label{EW-resonance}
		\tilde A_t(r_H)=\frac{\omega}{e},\qquad
		\tilde A_t(\infty)=0,
	\end{equation}
	so that the frequency is fixed by the horizon-to-infinity
	electrostatic potential difference,
	$e[\tilde A_t(r_H)-\tilde A_t(\infty)]=\omega$.
	This is the electroweak realization of the resonance mechanism
	encountered in nonlinear scalar $Q$-clouds
	\cite{Herdeiro:2020xmb}. Simultaneously,
	$A_0(r_H)=a_0(r_H)$ removes the temporal Higgs covariant derivative
	at the horizon. Although the spacetime metric is Schwarzschild, the
	cloud itself carries an electromagnetic field: only its gravitational
	backreaction is neglected. 
    {Also, at fixed horizon radius and frequency, the above boundary conditions also
		admit a condensate-free Coulomb solution,
        with $\phi=1$, $A_0=a_0= {4r r_H \omega}/{(r+r_H)^2}$, while other functions vanish.
        }.
	
	\subsection{Numerical results: branches, gaps and maximal horizon size}
	
	We solve the nonlinear matter equations with the same numerical
	strategy and dimensionless conventions used for the electroweak balls.
	The electric charge $\mathcal Q$ is read from the asymptotic
	flux (\ref{Qel}). The quantity denoted by $M$ in the figures is the
	energy of the matter fields outside the horizon, obtained by
	integrating their Killing energy density on a constant-$t$ slice;
	it should not be confused with the ADM mass of a fully backreacted
	BH spacetime.
	
\begin{figure}[h!]
		\centering
		\includegraphics[width=0.58\columnwidth]{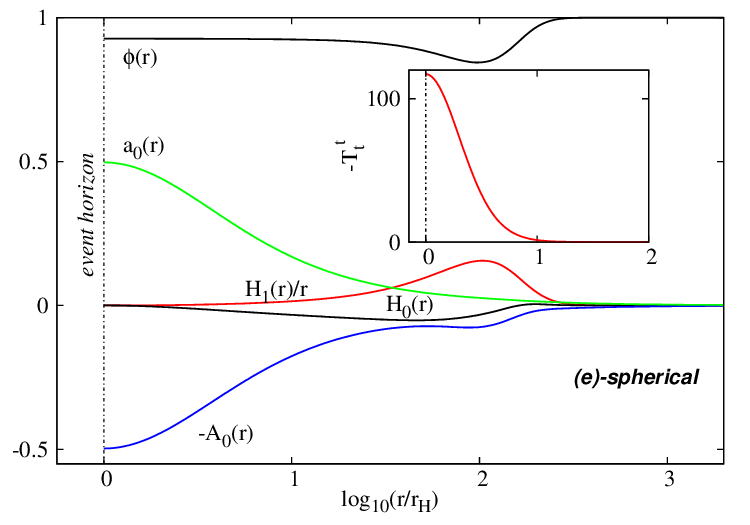}
        \captionsetup{font=small}
		\caption{Profiles of a representative spherical electric electroweak
			cloud on a Schwarzschild background, corresponding to the marked
			configuration in the left panel of Fig.~\ref{BH-Mw}.}
		\label{profileS}
	\end{figure}

\begin{figure*}[h!]
		\centering
		\includegraphics[width=0.48\textwidth]{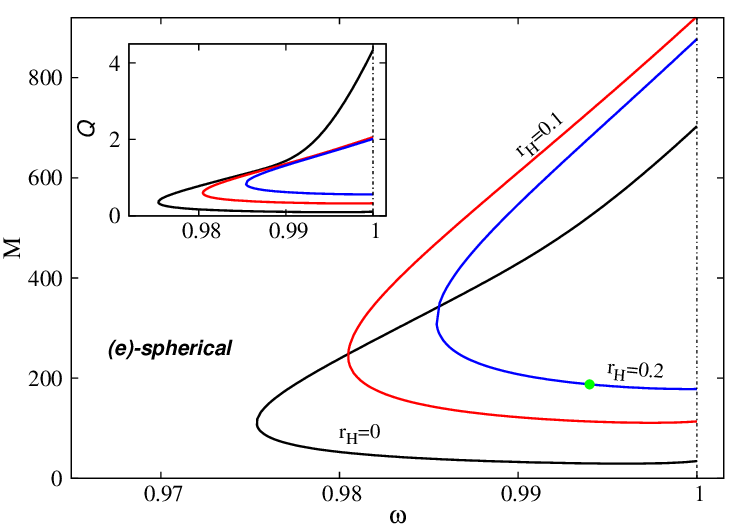}\hfill
		\includegraphics[width=0.48\textwidth]{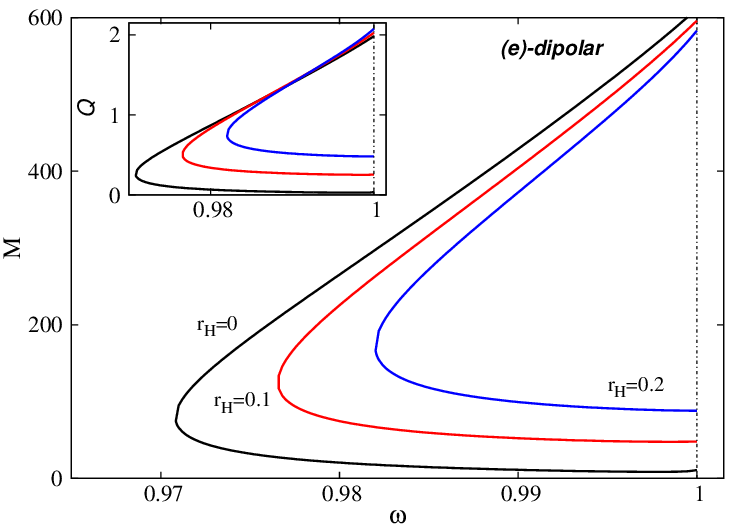}
        \captionsetup{font=small}
		\caption{Exterior matter energy $M$ and electric charge $\mathcal Q$
			as functions of the frequency $\omega$ for spherical (left) and
			dipolar (right) electric electroweak clouds on fixed Schwarzschild
			backgrounds. Here and in Fig.~\ref{BH-MAH},
			$g'/g=0.04$ and $\beta=0.109545$.}
		\label{BH-Mw}
	\end{figure*}

\begin{figure*}[h!]
		\centering
		\includegraphics[width=0.48\textwidth]{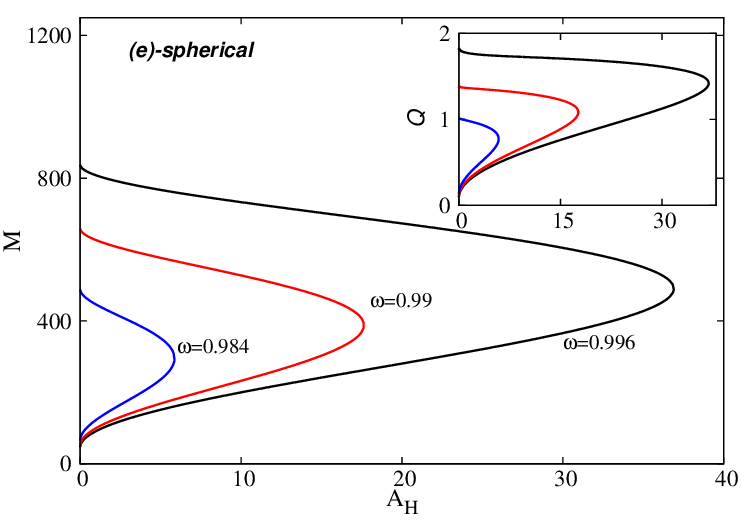}\hfill
		\includegraphics[width=0.48\textwidth]{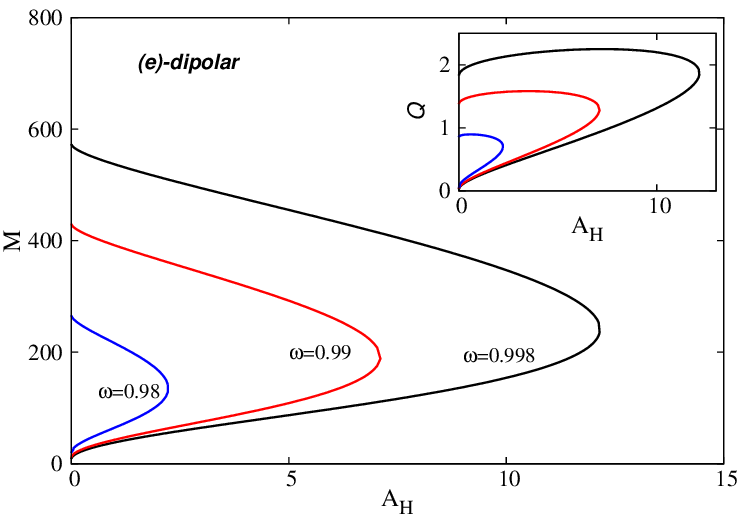}
        \captionsetup{font=small}
		\caption{Exterior matter energy $M$ and electric charge $\mathcal Q$
			as functions of the horizon area $A_H$ at fixed frequency, for
			spherical (left) and dipolar (right) electric electroweak clouds.
			Each family terminates at a maximal horizon area and approaches
			solitonic configurations as $A_H\to0$.}
		\label{BH-MAH}
	\end{figure*}

\begin{figure*}[h!]
		\centering
		\includegraphics[width=0.48\textwidth]{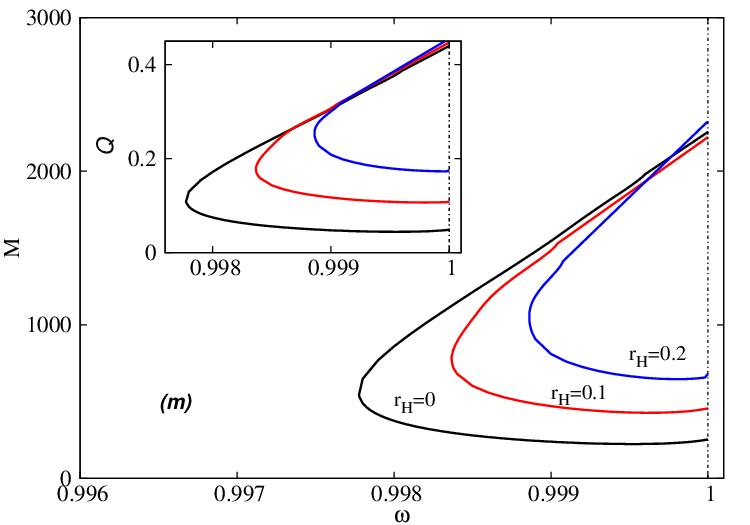}\hfill
		\includegraphics[width=0.48\textwidth]{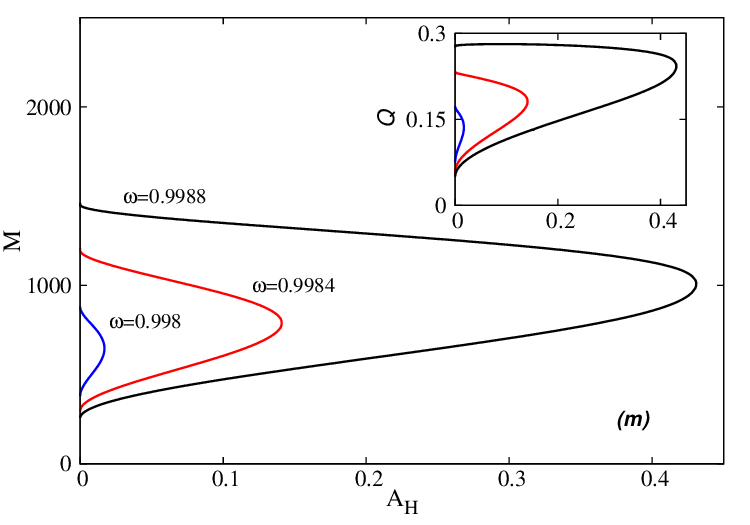}
        \captionsetup{font=small}
		\caption{Exterior matter energy $M$ and electric charge $\mathcal Q$
			for magnetic-type electroweak clouds as functions of frequency
			(left) and horizon area (right), for $g'/g=0.01$ and
			$\beta=0.1523$. The same characteristic branch structure is found
			as in the electric sectors.}
		\label{BH-Mw-m}
	\end{figure*}

	All three sectors investigated here, $i.e.$ electric spherical, electric
	dipolar and magnetic-type, admit localized configurations on the
	Schwarzschild background. Figure~\ref{profileS} illustrates a
	representative spherical electric cloud: the matter profiles are
	regular at the horizon and the energy density is concentrated in its
	vicinity. The more significant information, however, is the global
	structure of the families displayed in
	Figs.~\ref{BH-Mw}--\ref{BH-Mw-m}.
	
	At fixed horizon radius, varying $\omega$ yields two branches
	joining at a minimum frequency $\omega_{\min}(r_H)$; in the numerical
	families, this minimum increases with $r_H$.
	The same two-branch pattern occurs in the solitonic limit
	$r_H\to0$, and persists for both electric sectors
	(Fig.~\ref{BH-Mw}) and the magnetic sector
	(Fig.~\ref{BH-Mw-m}, left panel). Thus the Schwarzschild horizon
	deforms, rather than qualitatively changes, the characteristic
	frequency dependence of the parent electroweak balls.
	
	A complementary picture emerges when the frequency is held fixed
	and the horizon area is varied. The two branches meet at a maximal
	area $A_H^{\max}(\omega)$ and, as $A_H\to0$, approach the lower-
	and upper-branch electroweak balls, respectively
	(Figs.~\ref{BH-MAH} and \ref{BH-Mw-m}, right panel).
	In particular, the numerical families do not continue to
	arbitrarily large horizons. The maximal supported area generally
	increases with frequency over the explored ranges; its limiting
	behaviour as $\omega\to M_W$ remains to be established.
	
	A further distinctive feature is the absence of a small-amplitude
	limit along the families found numerically: both the exterior
	matter energy and the electric charge remain separated from zero.
	The clouds are therefore genuinely nonlinear continuations of
	electroweak balls, with no linear-cloud bifurcation  
    identified on
	these branches. 
   {In particular, no approach to the Coulomb solution with zero condensate amplitude
    was observed along the computed branches (in the spherically symmetric case this can be proved analytically).
    }
    A complete exclusion of other possible linear modes
	would require a separate spectral analysis.
	
	Finally, as in the solitonic case, our numerical searches have not
	produced cloud solutions for the simultaneously measured values of
	the electroweak couplings. The solutions established here thus
	exhibit the resonant gauge-Higgs mechanism in the broader parameter
	space of the bosonic WS model, rather than at the physical Standard
	Model point.

	\section{Conclusions}
	\label{section:conclusions}
	\label{final}
	
	We have presented numerical evidence for the existence of
	nonlinear electroweak clouds around Schwarzschild BHs in the
	bosonic Weinberg-Salam theory. These configurations are
	supported by harmonically time-dependent charged gauge fields,
	while their energy-momentum tensor remains static. Their
	compatibility with a regular event horizon relies on a
	resonance condition relating the field frequency to the horizon
	electrostatic potential. We have constructed electric-type
	spherical and dipolar clouds, as well as magnetic-type
	configurations, which continuously connect to electroweak
	balls \cite{Herdeiro:2026gbo} in the vanishing-horizon limit.
	
	Despite their different spatial structures, all families
	exhibit a remarkably similar behaviour. They possess two
	branches of solutions, exist only up to a maximal horizon
	size, and display a finite energy and electric charge gap,
	with no linear-cloud limit identified along the branches
	investigated. As a complementary result, we have constructed
	a new family of electric dipolar electroweak balls, extending
	the analogy with the multipolar Proca stars
	\cite{Herdeiro:2023wqf}.
	{
	Interestingly,  at fixed
	electric charge, these nonspherical solitons have
	lower energy than their spherical and magnetic counterparts \cite{Herdeiro:2026gbo}.
}
	
	Our results provide a new realization of the resonance
	mechanism \cite{Herdeiro:2020xmb}, showing how nonlinear
	condensates of massive charged vector fields can be supported
	outside a static BH horizon within a non-Abelian gauge-Higgs
	theory. Unlike the scalar case, the massive fields responsible
	for the condensate are gauge bosons whose masses originate
	from spontaneous symmetry breaking.
	
	The present solutions have been obtained in the probe limit
	and for electroweak couplings away from their measured values.
   {A contemporaneous study \cite{dosSantosCostaFilho:2026gwe} has established fully backreacted, spherically symmetric electrically charged black holes with resonant electroweak hair. It would be interesting to determine whether the dipolar electric and magnetic-type clouds constructed here likewise possess fully backreacted generalizations, and whether the probe families studied here can be continuously connected to the corresponding gravitating solutions as the gravitational coupling and matter parameters are varied.
   }
    Together with their dynamical
	stability, these questions provide natural directions for
	further investigation.

	\section*{Acknowledgement}
	%
	%
	This work is  supported by 
	CIDMA (\url{https://ror.org/05pm2mw36}) under the Portuguese Foundation for Science and Technology (FCT, \url{https://ror.org/ 00snfqn58}), Grants UID/04106/2025 (\url{https://doi.org/10.54499/UID/ 04106/2025}), UID/PRR/04106/2025    
	(\url{https://doi.org/10.54499/UID/PRR/ 04106/2025}), as well as the project Horizon Europe staff exchange (SE) programme HORIZON-MSCA2021-SE-01 Grant No.\ NewFunFiCO-101086251.

	
	\bibliographystyle{unsrt}
	
	\bibliography{biblio.bib}

\end{document}